\documentclass[12pt]{article}
\usepackage{epsf}

\begin{document}

\title{Forthcoming Close Angular Approaches of Planets to Radio
       Sources and Possibilities to Use Them as GR Tests}
\author{Z. M. Malkin, V. N. L'vov, and S. D. Tsekmeister \\
        Pulkovo Observatory, St. Petersburg, Russia \\
        e-mail: malkin@gao.spb.ru, epos@gao.spb.ru}
\date{Jul 22, 2009}
\maketitle

\begin{abstract}
During close angular approaches of solar system planets to astrometric radio sources, the apparent
positions of these sources shift due to relativistic effects and, thus, these events may be used for testing the theory
of general relativity; this fact was successfully demonstrated in the experiments on the measurements of
radio source position shifts during the approaches of Jupiter carried out in 1988 and 2002. An analysis, performed
within the frames of the present work, showed that when a source is observed near a planet's disk edge,
i.e., practically in the case of occultation, the current experimental accuracy makes it possible to measure the
relativistic effects for all planets. However, radio occultations are fairly rare events. At the same time, only Jupiter
and Saturn provide noticeable relativistic effects approaching the radio sources at angular distances of about
a few planet radii. Our analysis resulted in the creation of a catalog of forthcoming occultations and approaches
of planets to astrometric radio sources for the time period of 2008-2050, which can be used for planning experiments
on testing gravity theories and other purposes. For all events included in the catalog, the main relativistic
effects are calculated both for ground-based and space (Earth-Moon) interferometer baselines.
\end{abstract}
\bigskip

\section{Introduction}

Apart from the theory of general relativity (GR),
generally accepted among physicists and astronomers,
alternative theories of gravitation have been suggested.
The most serious of them do not contradict the available
observational data, but predict deviations from GR
for circumstances that have not been observed so far or
for accuracies higher than those that have been already
achieved. Therefore, the testing of gravity theories on
the basis of different methods and more and more precise
observations is an urgent astronomical and physical
problem.

Among the proposed GR tests are the Very Long Baseline
Interferometry (VLBI) observations of radio sources
at the time instants when solar system planets closely
approach them (Treuhaft and Lowe, 1991; Kopeikin,
2001; Fomalont and Kopeikin, 2003, 2008). In particular,
these papers present the observations of relativistic time
delays of radio source signals at the instances of close
approaches of Jupiter in 1988 and 2002. Similar observations
of Jupiter and Saturn are planned at Oleg Titov's proposal
by the International VLBI Service for Geodesy and
Astrometry (Schl\"uter and Behrend, 2007) in 2008-2009
(http://ivscc.gsfs.nasa.gov/program/opc.html). However,
the reduction of already available experimental data
showed that some formulated problems still have no
unambiguous answer, in particular, because the achieved
accuracy proved to be insufficient, mostly as a result of the
fairly large angular distances between the radio sources
which should be performed in the most favorable conditions,
first of all, during the closest approaches (further, we
imply apparent approaches) and occultations.

Note that so far the observations have been carried
out in an irregular way, i.e., an experimenter used the
closest in time approach of a planet to a known astrometric
radio source. Another researcher who decided to perform
a similar experiment has to search for forthcoming
approaches himself. In this situation, the most interesting
phenomena, i.e., the closest angular approaches, may be
missed. In addition, the observations of relativistic
effects caused by light deflections and gravitational time
delays near a planet require significant, usually international,
resources and should be planned well in advance.
Therefore, the possibility of early planning of the experiments
that can be performed in the most favorable conditions
is highly desirable.

In the present work, we tried to create a catalog of
approaches of solar system planets to astrometric radio
sources for the convenient and reliable planning of
observations. During the work, we found that two characteristics
of such experiments, which are currently
treated as obvious, are wrong. First, we found that such
events occur much more often than it had been considered
before. Second, one of the most interesting relativistic
effects related to the speed of gravity propagation
proved to be observable not only in the case of giant
planets, but in the case of all other planets, including
Pluto. In addition, the development of space technology
makes it possible to perform VLBI experiments with
baselines of several hundred thousand kilometers; with
these baselines, the main relativistic effects can be
observed in the case of all planets, including the largest
among the minor planets.

\section{Observable effects}

When a planet approaches a radio source, one can
observe two relativistic effects which contribute into
the measured interferometric delay, defined as the difference
between the instants of electromagnetic wave
front arrivals to two VLBI antennas. These effects are
the Shapiro time delay $\Delta$ and the delay of gravity propagation
$\Delta_P$ caused by the ray propagation near a moving
body, i.e., near a planet in our case. Kopeikin (2001)
showed that their measurement makes it possible to
determine two fundamental relativistic parameters: $\gamma$,
which is equal to unity in GR, and the parameter of
gravity propagation $\delta$, which is equal to zero in GR, i.e.,
the gravity speed is considered to be equal to the speed
of light. The values of these effects can be estimated
from the formulas derived by Kopeikin (2001; Eq. 13),
which after the obvious transformations take the form
\begin{equation}
\Delta \cong \frac{2(1+\gamma)GMrB}{c^3Rd} \,, \quad \Delta_P \cong (1+\delta)\frac{\Delta v}{cd} \,,
\end{equation}
where $GM$ is the planetocentric gravitational constant,
$B$ is the interferometer baseline, $r$ is the apparent angular
radius of the planet, $d$ is the angular distance
between the source and the planet center for a station,
$R$ is the planet radius, $v$ is its orbital speed, and $c$ is the
speed of light. One can see that $\Delta$ and $\Delta_P$ achieve their
maxima at the edge of the planet's disk, i.e., when $d=r$.
In this case, Eq. (1) within the frames of GR can be rewritten
in the form
\begin{equation}
\Delta \cong \frac{4GMB}{c^3R} \,, \quad \Delta_P \cong \frac{\Delta v}{cr} \,.
\end{equation}

Equation (2) shows that at the edge of the disk, $\Delta$
does not depend on the apparent size of the disk, i.e., on
the planet distance, and $\Delta_P$ reaches its maximum when
the distance between the Earth and the planet becomes
the largest. The maximum values of the observable relativistic
effects are present in Table~1 for ground-based
and space-based (Earth-Moon) interferometer baselines.
Taking into account the linear dependence of the
considered effects on B, these values can be easily
recalculated for any baseline.

Note that the relativistic effects that are of the order
of nanoseconds  can be measured using the simplest interferometric
technique of the group delay measurements. At the
same time, the effects of the order of picoseconds require differential
phase measurements in carefully planned observations.
One can reasonably expect a significant
increase of the accuracy of relativistic effect measurements
with setting in operation the new VLBI 2010 stations (Behrend et al., 2008).

Calculations show that nearly all planets during
their closest approaches to radio sources provide relativistic
effects, especially for , that can be observed
even with ground-based interferometers. However,
with an increase of the angular distances between the
planets' centers and radio sources to 30 arcsec, the relativistic
effects become so weak that ground-based
interferometers can produce sufficient accuracy only
for Jupiter and Saturn (Table~2).

\section{Precomputation of occultations and approaches}

On the basis of the results presented in the previous
section, one can conclude that the instants of occultations
should be precomputated for all planets from
Venus to Neptune, and for Jupiter and Saturn the circumstances
of the approaches should be precomputated
in more detail. In this case, all of the approaches of
planets to radio sources that provide the most noticeable
relativistic effects of signal propagation will be
taken into account. Their amount makes it possible to
plan experiments not involving Mercury, Mars, and
minor planets, for these planets' relativistic effects can
be measured, but with fairly large relative errors. However,
the data omitted in this paper can be easily supplied
by the authors at the request of interested parties.

Most computations of the circumstances of planet
approaches to radio sources were performed using the
codes APPROACH and OCCULT, which utilize the Ephemeride
Package for Objects of the Solar System (EPOS;
L'vov et al., 2001) data and environment. Source coordinates
were taken from the Goddard center of space
flight's catalog (Petrov, 2008), adding sources from the
ICRF-2 catalog (Fey et al., 2004). The total number of
sources proved to be 3958; their list and optical characteristics
are available at the website
http://www.gao.spb.ru/english/as/ac\_vlbi/sou\_car.dat.

The list of source occultations by planets is presented
in Table 3, and the circumstances of Jupiter and
Saturn approaches to radio sources are given in Tables
4 and 5, respectively. The tables exhibit the circumstances
of all occultations and approaches closer than
10. over the time interval from September 2008 (the
time of writing this paper) to 2050. There is no occultation
for Uranus and Neptune over this time interval. An
interesting feature of the list is the presence of multiple
approaches due to the retrograde apparent motion of the
planets. In this case, a planet approaches a radio source
from different directions, which may have a particular
interest for studying the influence of a moving planet on
the signal delay ( term).

The apparent angular diameters of planets in Tables
4 and 5 are calculated from their mean radii. The $\Delta$ and
$\Delta_P$ values are calculated for a 8000-km long baseline. In
this case, we used an equation which obviously follows
from Eq. 1 (GR case)
\begin{equation}
\Delta \cong \frac{4GMB}{c^3Dd} \,,
\end{equation}
where $D$ is the distance between the Earth and the
planet. Data for other baselines, including space-based
ones, can be easily obtained by proportionally recalculating
the present values. The results of the experiment
performed in 2002 (Fomalont and Kopeikin, 2003)
showed that the relativistic time delay  $\Delta_P$ = 6 ps (more
precisely, an equivalent light deflection of 51 $\mu$s)
proved to be measurable with an accuracy of 20\% using
the VLBA interferometer supplemented by the 100-meter antenna
at Effelsberg, Germany. Less close
approaches should be observed with space interferometers
for which the effect is larger proportionally to the
ratio of space/ground-based interferometer baseline
lengths.

Note that in the present work, the approach circumstances
are calculated for the geocenter. For a real
observer, the angular distances will differ from the values
presented in Tables 4 and 5 up to the value of the
$R_0/D$ ratio, where $R_0$ is the distance from the geocenter
to the mid-baseline, depending on the position angle
and baseline orientation. Clearly, this difference
reaches its maximum at the epochs of oppositions, and
for ground-based interferometers may be as large as $3''$
for Jupiter and $1''$ for Saturn. The data presented in
Table 3 were calculated for ground-based observations;
in regard to space interferometers, the relevant calculations
should be performed for their specific configurations.

\section{Conclusions}

We have calculated the circumstances of the
approaches of solar system planets to astrometric radio
sources for the time interval of 2008-2050. Especially
interesting are the radio occultations for which the relativistic
effects reach their maximum values. These
occultations make it possible to measure the effects
with minimum relative errors, which is a question of the
utmost importance for testing gravity theories. One can
efficiently use all of the planets from Venus to Saturn.

The present work demonstrates that the apparent
approaches of planets to radio sources and even the
radio occultations are not as rare of events as it is generally
considered. The number of events in consideration grows with
the expansion of the list of ecliptic
radio sources, thus increasing the possibilities to perform
relevant experiments.

\section{References}

%\leftskip=5mm
%\parindent=-5mm
\parindent=0mm
\parskip = 1em

Behrend, D, Boehm, J, Charlot, P, et al., Proc. 2007 IAG
General Assembly. Observing our Changing Earth.
Perugia, Italy, July 2-13, 2007, pp. 833-840.

Fey, A.L., Ma C., Arias E.F., et al. The second extension of
the International Celestial Reference Frame: ICRF-Ext.2, Astron. J., vol. 127, pp. 3587-3608.

Fomalont, E.B. and Kopeikin, S.M., The Measurement of the
Light Deflection from Jupiter: Experimental Results,
Astrophys. J., 2003, vol. 598, pp. 704-711.

Fomalont, E.B and Kopeikin, S.M, Radio interferometric
tests of general relativity, Proc. IAU Symposium
No. 248A ``Giant Step: from Milli- to Micro-arcsecond
Astrometry'', 2008, pp. 383-386.

Kopeikin, S.M., Testing the Relativistic Effect of the Propagation
of Gravity by Very Long Baseline Interferometry,
Astrophys. J., 2001, vol. 556, pp. L1-L5.

L'vov, V.N., Smekhacheva, R.I., and Tsekmeister, S.D.,
EPOS - programmnaya sistema dlya podderzhki issledovanii
ob'ektov Solnechnoi sistemy, Trudy Konf.
``Okolozemnaya Astronomiya'' (Proc. Conf. ``Near-Earth
Astronomy''), Zvenigorod, 2001, pp. 21-25.

Petrov, L., Goddard VLBI astrometric catalogue 2008b. \\
http://vlbi.gsfc.nasa.gov/solutions/2008b\_astro.

Schluter, W. and Behrend, D., The International VLBI Service
for Geodesy and Astrometry (IVS): Current Capabilities
and Future Prospects, J. Geodesy, 2007, vol. 81,
pp. 379-387.

Treuhaft, R.N. and Lowe, S.T., A Measurement of Planetary
Relativistic Deflection, Astron. J., 1991, vol. 102,
pp. 1879-1888.

\eject
%%%%%%%%%%%%%%%%%%%%%%%%%%%%%%%%%%%%%%%%%%%%%%%%%%%%%%%%%%%%%%%%%%%

\begin{table}
\centering
\small
\caption{The maximum values of relativistic effects for the case of observations at the edges of a planet's disks, ns}
\begin{tabular}{c|c|c|c|c|c|c|c|c|c}
\hline
Baseline, & Effect& Mercury& Venus& Mars& Jupiter& Saturn& Uranus& Neptune& Pluto \\
thousand km &&&&&&&& \\
\hline
8   & $\Delta$  & 0.01& 0.06& 0.02& 2.1& 0.77& 0.27& 0.33& 0.00 \\
    & $\Delta_P$& 0.15& 0.32& 0.13& 1.2& 0.68& 0.73& 1.14& 0.09 \\
400 & $\Delta$  & 0.5 & 3.2 & 0.75& 110& 39  & 14  & 16  & 0.05 \\
    & $\Delta_P$& 7.3 & 16  & 6.7 & 62 & 34  & 37  & 57  & 4.3  \\
\hline
\end{tabular}
\end{table}

%%%%%%%%%%%%%%%%%%%%%%%%%%%%%%%%%%%%%%%%%%%%%%%%%%%%%%%%%%%%%%%%%%%

\begin{table}
\centering
\small
\caption{The values of relativistic effects for a 30-arcsec angular distance between the planet center and the source, ns}
\begin{tabular}{c|c|c|c|c|c|c|c|c|c}
\hline
Baseline, & Effect& Mercury& Venus& Mars& Jupiter& Saturn& Uranus& Neptune& Pluto \\
thousand km &&&&&&&& \\
\hline
8   & $\Delta$  & 0.00& 0.01& 0.00& 1.1 & 0.20& 0.02& 0.01& 0.00 \\
    & $\Delta_P$& 0.00& 0.01& 0.00& 0.33& 0.04& 0.00& 0.00& 0.00 \\
400 & $\Delta$  & 0.04& 0.51& 0.05& 56  & 9.8 & 0.78& 0.60& 0.00 \\
    & $\Delta_P$& 0.05& 0.41& 0.03& 17  & 2.2 & 0.12& 0.08& 0.00 \\
\hline
\end{tabular}
\end{table}

%%%%%%%%%%%%%%%%%%%%%%%%%%%%%%%%%%%%%%%%%%%%%%%%%%%%%%%%%%%%%%%%%%%

\begin{table}
\centering
\small
\caption{Occultations of astrometric radio sources by planets}
\begin{tabular}{l|c|c|r|r|l}
\hline
Planet & Date, y m d & Source & \multicolumn{2}{|c|}{$\alpha$ and $\delta$, J2000} & Notes, seeing \\
\cline{4-5}
&&& h~~m~~s~ & d~~m~~s~ & \\
\hline
Venus   & 2011 02 26.6 & 1946--200 & 19 49 53 & --19 57 13 & S. America, Australia, Antarctica    \\
Mars    & 2011 05 03.8 & 0127+084  &  1 30 28 &   +8 42 46 & America                              \\
Venus   & 2012 12 24.4 & 1631--208 & 16 34 30 & --20 58 26 & Africa, S. America, Antarctica       \\
Venus   & 2015 08 06.8 & 0947+064  &  9 50 03 &   +6 15 04 & America                              \\
Venus   & 2020 01 16.7 & 2220--119 & 22 22 56 & --11 44 26 & Europe, Africa, S. America           \\
Venus   & 2020 07 17.7 & 0446+178  &  4 49 13 &  +17 54 32 & America                              \\
Jupiter & 2025 09 18.6 & 0725+219  &  7 28 21 &  +21 53 06 & America, Antarctica                  \\
Saturn  & 2028 10 24.8 & 0223+113  &  2 25 42 &  +11 34 25 & Annular; Asia, Africa, Europe        \\
Jupiter & 2033 02 04.2 & 2104--173 & 21 07 27 & --17 08 10 & S. America, Australia, Antarctica    \\
Venus   & 2035 07 03.3 & 0558+234  &  6 01 47 &  +23 24 53 & Europe, Asia, Africa                 \\
Venus   & 2037 01 03.8 & 1734--228 & 17 37 02 & --22 51 55 & S. America, Australia, Antarctica    \\
Jupiter & 2043 02 01.1 & 1734--228 & 17 37 02 & --22 51 55 & Australia, Asia, Africa, Antarctica  \\
Venus   & 2043 02 15.6 & 1858--212 & 19 01 04 & --21 12 01 & America                              \\
Venus   & 2043 02 17.7 & 1908--211 & 19 11 54 & --21 02 44 & America, Australia                   \\
Jupiter & 2045 09 24.4 & 2221--116 & 22 24 08 & --11 26 21 & S. America, Australia, Antarctica    \\
Venus   & 2049 01 13.5 & 2243--081 & 22 45 49 &  --7 55 19 & Asia, Europe, Africa                 \\
Venus   & 2049 11 02.2 & 1333--082 & 13 36 08 &  --8 29 52 & Africa, Antarctica                   \\
\hline
\end{tabular}
\end{table}

%%%%%%%%%%%%%%%%%%%%%%%%%%%%%%%%%%%%%%%%%%%%%%%%%%%%%%%%%%%%%%%%%%%

\begin{table}
\centering
\small
\caption{Apparent approaches of Jupiter to astrometric radio sources}
\def\arraystretch{0.95}
\begin{tabular}{l|c|r|r|c|c|c|c}
\hline
Date, y m d & Source & \multicolumn{2}{|c|}{$\alpha$ and $\delta$, J2000} & d, & r, & $\Delta$, & $\Delta_P$, \\
\cline{3-4}
&& h~~m~~s~ & d~~m~~s~ & arcsec & arcsec & ps & ps \\
\hline
2008 11 19.0& 1922--224&  19 25 40&  --22 19 35&   83&  17&  443 & 48 \\
2009 03 08.6& 2104--173&  21 07 27&  --17 08 10&  277&  16&  127 & 4.1 \\
2011 07 03.6& 0210+119 &   2 13 05&   +12 13 11&  341&  18&  116 & 3.0 \\
2011 09 13.1& 0229+131 &   2 31 46&   +13 22 55&  149&  23&  328 & 20 \\
2012 02 04.0& 0201+113 &   2 03 47&   +11 34 45&  490&  19&  83  & 1.5 \\
2012 02 20.3& 0210+119 &   2 13 05&   +12 13 11&  342&  18&  114 & 3.0 \\
2013 02 28.1& 0420+210 &   4 23 02&   +21 08 02&  216&  19&  191 & 8.0 \\
2013 10 23.0& 0723+219 &   7 26 14&   +21 53 20&  123&  20&  343 & 25 \\
2013 11 07.0& 0725+219 &   7 28 21&   +21 53 06&  388&  21&  114 & 2.6 \\
2013 11 22.1& 0723+219 &   7 26 14&   +21 53 20&  351&  21&  132 & 3.4 \\
2017 10 13.7& 1352--104&  13 54 47&  --10 41 03&   69&  15&  471 & 62 \\
2019 10 28.4& 1723--229&  17 26 59&  --22 58 02&  184&  16&  192 & 9.4 \\
2020 08 02.0& 1922--224&  19 25 40&  --22 19 35&  79 &  23&  631 & 72 \\
2020 10 24.2& 1922--224&  19 25 40&  --22 19 35&  355&  18&  112 & 2.8 \\
2021 02 19.9& 2104--173&  21 07 27&  --17 08 10&  149&  16&  232 & 14 \\
2022 11 13.8& 2354--021&  23 57 25&   --1 52 16&  159&  22&  304 & 17 \\
2022 12 04.1& 2354--021&  23 57 25&   --1 52 16&  177&  21&  257 & 13 \\
2023 06 11.1& 0210+119 &   2 13 05&   +12 13 11&   28&  17&  1321&  426 \\
2023 11 05.4& 0229+131 &   2 31 46&   +13 22 55&  199&  24&  261 & 12 \\
2024 01 02.1& 0210+119 &   2 13 05&   +12 13 11&  396&  21&  117 & 2.6 \\
2025 09 15.4& 0723+219 &   7 26 14&   +21 53 20&  215&  17&  174 & 7.3 \\
2025 10 25.0& 0741+214 &   7 44 47&   +21 20 00&   30&  19&  1374&  406 \\
2025 11 29.1& 0741+214 &   7 44 47&   +21 20 00&  274&  22&  169 & 5.5 \\
2029 03 15.3& 1333--082&  13 36 08&   --8 29 52&  432&  21&  105 & 2.2 \\
2029 09 28.5& 1352--104&  13 54 47&  --10 41 03&   47&  15&  704 & 136 \\
2031 02 23.2& 1734--228&  17 37 02&  --22 51 55&  261&  17&  142 & 4.9 \\
2031 06 07.1& 1734--228&  17 37 02&  --22 51 55&   55&  23&  877 & 143 \\
2031 10 05.6& 1723--229&  17 26 59&  --22 58 02&  312&  18&  121 & 3.5 \\
2033 02 27.2& 2126--158&  21 29 12&  --15 38 41&  417&  16&  83  & 1.8 \\
2034 01 28.9& 2245--091&  22 47 52&   --8 50 22&  342&  17&  105 & 2.8 \\
2035 05 14.0& 0201+113 &   2 03 47&   +11 34 45&  433&  16&  81  & 1.7 \\
2035 05 24.1& 0210+119 &   2 13 05&   +12 13 11&  173&  17&  206 & 11 \\
2037 05 28.4& 0558+234 &   6 01 47&   +23 24 53&  306&  16&  112 & 3.3 \\
2037 08 27.9& 0725+219 &   7 28 21&   +21 53 06&  159&  16&  222 & 13 \\
2037 09 19.0& 0741+214 &   7 44 47&   +21 20 00&   29&  17&  1271&  391 \\
2041 09 11.6& 1352--104&  13 54 47&  --10 41 03&   74&  16&  455 & 55 \\
2045 01 20.1& 2104--173&  21 07 27&  --17 08 10&  192&  16&  180 & 8.4 \\
2045 02 12.0& 2126--158&  21 29 12&  --15 38 41&  283&  16&  121 & 3.8 \\
2045 05 29.4& 2245--091&  22 47 52&   --8 50 22&  459&  19&  91  & 1.8 \\
2045 09 20.3& 2223--114&  22 25 44&  --11 13 41&  228&  24&  224 & 8.9 \\
2045 12 04.5& 2223--114&  22 25 44&  --11 13 41&  466&  19&  89  & 1.7 \\
2046 01 10.7& 2245--091&  22 47 52&   --8 50 22&   83&  17&  449 & 48 \\
2047 04 28.4& 0201+113 &   2 03 47&   +11 34 45&  294&  16&  118 & 3.6 \\
2047 05 08.3& 0210+119 &   2 13 05&   +12 13 11&  308&  16&  113 & 3.3 \\
2049 05 11.4& 0558+234 &   6 01 47&   +23 24 53&  129&  16&  272 & 19 \\
2049 08 29.5& 0741+214 &   7 44 47&   +21 20 00&  179&  16&  196 & 9.9 \\
\hline
\end{tabular}
\end{table}

%%%%%%%%%%%%%%%%%%%%%%%%%%%%%%%%%%%%%%%%%%%%%%%%%%%%%%%%%%%%%%%%%%%

\begin{table}
\centering
\small
\caption{Apparent approaches of Saturn to astrometric radio sources}
\begin{tabular}{l|c|r|r|c|c|c|c}
\hline
Date, y m d & Source & \multicolumn{2}{|c|}{$\alpha$ and $\delta$, J2000} & d, & r, & $\Delta$, & $\Delta_P$, \\
\cline{3-4}
&& h~~m~~s~ & d~~m~~s~ & arcsec & arcsec & ps & ps \\
\hline
2009 02 10.2& 1125+062 &  11 27 37&     +5 55 32&  80 &  9 &   92&  7.7 \\
2009 06 26.0& 1109+076 &  11 12 10&     +7 24 49&  146&  8 &   44&  2.0 \\
2015 06 19.1& 1548--177&  15 51 15&   --17 55 02&  156&  9 &   44&  1.9 \\
2015 11 19.1& 1614--195&  16 17 27&   --19 41 32&  64 &  7 &   88&  9.1 \\
2016 11 22.9& 1658--217&  17 02 10&   --21 30 03&  194&  7 &   29&  1.0 \\
2017 12 13.3& 1752--225&  17 55 26&   --22 32 11&  73 &  7 &   78&  7.1 \\
2021 08 10.8& 2044--188&  20 47 38&   --18 41 41&  20 &  9 &  347&  115 \\
2021 12 08.1& 2044--188&  20 47 38&   --18 41 41&  114&  8 &   52&  3 \\
2023 04 13.3& 2221--116&  22 24 08&   --11 26 21&  33 &  8 &  183&  37 \\
2023 04 18.2& 2223--114&  22 25 44&   --11 13 41&  276&  8 &   22&  0.5 \\
2024 01 04.6& 0220--119&   2 13 05&    +12 13 11&  370&  8 &   16&  0.3 \\
2024 03 18.5& 2252--090&  22 55 04&   --08 44 04&  158&  8 &   37&  1.5 \\
2026 04 01.5& 0019--001&   0 22 25&     +0 14 56&  472&  8 &   13&  0.2 \\
2026 10 19.0& 0037+011 &   0 40 14&     +1 25 46&  145&  9 &   51&  2.3 \\
2028 05 20.6& 0208+106 &   2 11 13&    +10 51 35&  79 &  8 &   77&  6.5 \\
2030 11 31.0& 0409+188 &   4 12 46&    +18 56 37&  306&  10&   25&  0.5 \\
2032 04 03.5& 0503+216 &   5 06 34&    +21 41 00&  71 &  9 &   93&  8.8 \\
2033 05 24.2& 0620+227 &   6 23 18&    +22 41 36&  206&  8 &   31&  1.0 \\
2034 06 15.7& 0725+219 &   7 28 21&    +21 53 06&  38 &  8 &  163&  28 \\
2034 07 16.2& 0741+214 &   7 44 47&    +21 20 00&  157&  8 &   39&  1.7 \\
2037 01 16.1& 1013+127 &  10 15 44&    +12 27 07&  72 &  10&  103&  9.6 \\
2037 07 24.1& 1013+127 &  10 15 44&    +12 27 07&  233&  8 &   26&  0.8 \\
2043 10 18.4& 1459--149&  15 02 25&   --15 08 53&  220&  7 &   26&  0.8 \\
2044 02 27.6& 1548--177&  15 51 15&   --17 55 02&  33 &  8 &  193&  39 \\
2045 09 20.4& 1614--195&  16 17 27&   --19 41 32&  46 &  8 &  132&  19 \\
2046 09 17.5& 1658--217&  17 02 10&   --21 30 03&  51 &  8 &  120&  16 \\
2047 10 17.1& 1752--225&  17 55 26&   --22 32 11&  367&  8 &   16&  0.3 \\
2048 11 28.4& 1853--226&  18 56 36&   --22 36 17&  321&  7 &   18&  0.4 \\
\hline
\end{tabular}
\end{table}

\end{document}